 \documentclass[11pt,twocolumn]{article}

 \pdfoutput=1

\usepackage[round]{natbib}   

\usepackage[english]{babel}

\usepackage{graphicx}
\usepackage{amsmath}

\usepackage{enumitem}

\usepackage{float}
\usepackage{ifthen}
\usepackage{hyperref}

\usepackage{titlesec}

\titlespacing*{\section}{0pt}{1.1\baselineskip}{\baselineskip}

\begin{document}



\twocolumn[
\begin{center}
{\bf \Large 
Comments on  ``MSE minus CAPE is the True Conserved Variable}\\
\vspace*{2mm}
{\bf \Large 
for an Adiabatically Lifted Parcel''.
}\\
\vspace*{3mm}
{\Large by Pascal Marquet}. \\
\vspace*{3mm}
{\large  CNRM/GMAP-GAME, M\'et\'eo-France / CNRS UMR3589.
 Toulouse. France.}
\\ \vspace*{2mm}
{\large  \it E-mail: pascal.marquet@meteo.fr}
\\ \vspace*{2mm}
{\large     Submitted to the}
{\large \it Journal of Atmospheric Science}
{\large (30 September 2015, revised 4 December 2015).} \\
\vspace*{1mm}
\end{center}
]





 \section{Introduction} 
\vspace*{-5mm}

In a recent paper, \citet[hereafter R15]{Romps2015} argues that the quantity ``MSE$\: - \:$CAPE'' must be used as a true conserved variable for an adiabatically lifted parcel, where MSE is the moist-air static energy and ``CAPE'' is expected to be the so-called convective available energy.

It is shown in this comment that the quantity denoted by CAPE in R15 is the opposite of the convective available energy.
It is explained that the vertical adiabatic ascent considered in R15 is not realistic, since it generates condensed water of the order of $10$ to $20$~g~kg${}^{-1}$ at height above $6$~km.
Moreover, the thermodynamic equations are written in R15 by making several assumptions, not all of which are explicitly mentioned.

This comment aims to clarify the hypotheses made in R15.
It will show that these assumptions call into question the validity of the moist-air internal energy, enthalpy and entropy functions in R15.
It also demonstrates that it is possible to obtain more precise and general formulations for moist-air energy, enthalpy and entropy functions, in particular by using the third law of thermodynamics.
The large differences between the thermodynamics formulas derived in R15 and those depending on the third law are illustrated by studying a realistic pseudo-adiabatic vertical profile.

The same notations as in R15 will be used as far as possible in this comment.

\vspace*{-3mm}

 \section{The convective available energy} 
\vspace*{-5mm}

The convective available potential energy CAPE$\, (z)$ is defined in R15 by the vertical integral of the parcel's buoyancy $b = g \: ( \rho_e / \rho - 1 )$ between the height $z$ to some fixed reference height $z_{\rm \,  top}$.
This integral decreases with height if $b \approx g \: ( T - T_e )/ T_e$ is positive, leading to a wrong definition of the CAPE.

The CAPE must be computed by integrating the parcel's buoyancy from the height of the level of free convection (LFC, at the surface in R15) to the height of neutral buoyancy (LBN, at the top of the vertical profile in R15).
The CAPE for a parcel ascending from the LFC is thus defined at a certain height $z>z_{\rm \:  LFC}$ by 
\vspace*{-5mm}
\begin{align}
 \mbox{CAPE}\:(z) & \: = \; 
   \int_{z_{\rm \:  LFC}}^{z} b(z') \; dz'
   \: , \label{eq_CAPE_true}
\end{align}
leading to $\partial\mbox{CAPE}/\partial z =  b(z)$.
This definition increases with height if $b$ is positive.

The definition (\ref{eq_CAPE_true}) is retained in Eq.~(10) in \citet{RiehlMalkus58}, where it is explained that $\int b \: dz \;$ ``{\it measures the vertical kinetic energy acquirable during ascent from parcel method calculations\/}'' and where it is shown that MSE$\, + \int b \: dz $ is a constant for adiabatic motions.

The quantity conserved in both Eq.~(10) in \citet{RiehlMalkus58} and Eqs.~(3) and (7) in R15 is thus the sum ``MSE$(z) \: + \: $CAPE$(z)$''.
As a consequence, the title of R15 should begin with ``MSE {\it plus\/} CAPE is the True Conserved Variable...'', with the CAPE defined by (\ref{eq_CAPE_true}) which increases with height.

\vspace*{-2mm}

 \section{The Bernoulli equation - Non-hydrostatic effects} 
\vspace*{-5mm}

There is a close link between the kinetic energy of the vertical wind KE$\: = w^2(z)/2$ and the CAPE defined by (\ref{eq_CAPE_true}).
This link is explicitly described in Eq.~(8) in \citet{MaddenRobitaille70}, where it is explained that KE$\: \approx w_0^2/2 + \mbox{CAPE}\:(z)$.
This corresponds to the Bernoulli vision considered in Eq.~(5) in \citet{RiehlMalkus57}, Eq.~(7) in \citet{MaddenRobitaille70} and Eq.~(7) and after Eq.~(12) in \citet{Betts74}.

The Bernoulli equation states that the conserved quantity is of the form $h+ g \: z + w^2/2$, with  MSE and the CAPE replaced by $h+ g \: z $ and $w^2/2$, respectively.
It might thus be possible to use a Bernoulli function to derive an alternative vision of the approach described in R15.
However, it is explained (end of section~2 in R15) that CAPE may not be converted into KE, but, instead, dissipated into environmental turbulence and wave energy.
For these reasons, MSE$\: + \:$KE would not be conserved in adiabatic motions.

It is assumed in several places in R15 that $p = p_e$, on the one hand, and that there may be significant pressure perturbation $p' = p - p_e$ and non-hydrostatic effects, in the other hand.
These assumptions seem inconsistent and it is difficult to appreciate the impact of this contradiction on the results derived in R15, including the conservation of ``MSE$\: + \:$CAPE'' or the non-conversion of CAPE into KE.



\vspace*{-2mm}

 \section{The moist-air internal energy and enthalpy} 
\vspace*{-5mm}

The first law of thermodynamics is written in Eq.~(1) of R15 in terms of a quantity denoted by $E_i$ in this comment, leading to
\vspace*{-2mm}
\begin{equation}
 E_i \; = \; c_{vm} \: (T - T_{\rm trip}) 
   \: + \: q_v \: E_{0v}
   \: - \: q_s \: E_{0s}
   \: , \label{eq_Ei_star}
\end{equation}
where 
$c_{vm} = q_a \: c_{va} + q_v \: c_{vv} +q_l \: c_{vl} + q_s \: c_{vs}$ 
is the heat capacity at constant volume for moist air and
$T_{\rm trip}$ the triple-point temperature.

It is suggested in R15 that $E_i$ given by (\ref{eq_Ei_star}) is the general moist-air ``specific internal energy'', with no mention made to the hypotheses required to established (\ref{eq_Ei_star}).
It is shown in this section that it is only valid for adiabatic motions of a closed parcel of moist air.

It is also assumed in R15 that ``the constant $E_{0v}$ is the difference in specific internal energy between water vapor and liquid at the triple-point temperature'' and that
``$E_{0s}$ is the difference in specific internal energy between water liquid and solid at the triple-point temperature.''
This means that $E_{0v} = e_{iv0}-e_{il0}$ and $E_{0s} = e_{il0}-e_{is0}$, where $e_{iv0}$, $e_{il0}$, $e_{il0}$ and $e_{is0}$ are the specific reference internal energies at $T = T_{\rm trip}$.
It is shown in this section that the true moist-air specific internal energy $e_i$ is not equal to $E_i$ given by (\ref{eq_Ei_star}).

Following the method described in 
\citet[hereafter M15]{Marquet15a} and \citet[hereafter MG15]{Marquet_Geleyn15b}
the moist-air internal energy is defined by
\vspace*{-6mm}
\begin{align}
 e_i & = \;    q_a \: e_{ia} 
    \: + \: q_v \: e_{iv} 
    \: + \: q_l \: e_{il} 
    \: + \: q_s \: e_{is} 
    \: . \label{eq_Ei} 
\end{align}
Internal energies of dry air and water species can be computed by assuming that all heat capacities at constant volume are constant in the atmospheric range of temperature, leading to
\vspace*{-3mm}
\begin{align}
 e_{ia} & = \; c_{va} \: (T - T_{\rm trip}) 
    \: + \: e_{ia0}
    \: , \label{eq_Eia} \\
 e_{iv} & = \: c_{vv} \: (T - T_{\rm trip}) 
    \: + \: e_{iv0}
    \: , \label{eq_Eiv} \\
 e_{il} & = \: c_{vl} \: (T - T_{\rm trip}) 
    \: + \: e_{il0}
    \: , \label{eq_Eil} \\
 e_{is} & = \: c_{vs} \: (T - T_{\rm trip}) 
    \: + \: e_{is0}
    \: . \label{eq_Eis}
\end{align}
The reference values of internal energies $e_{ia0}$ to $e_{is0}$ are computed at the triple-point temperature, as in R15.
If ($\ref{eq_Eia}$)-($\ref{eq_Eis}$) are inserted into ($\ref{eq_Ei}$) and $q_a = 1 - q_t$ is taken into account, where $q_t = q_v + q_l + q_s$ is the total water content, the moist-air specific internal energy can be written as
\vspace*{-3mm}
\begin{align}
 e_i & = \; c_{vm} \: (T - T_{\rm trip}) 
   \: + \: q_v \: (e_{iv0}-e_{il0})
   \: - \: q_s \: (e_{il0}-e_{is0})
  \nonumber \\
  & \quad \: + \: q_t \: (e_{il0}-e_{ia0}) \: + \: e_{ia0}
   \: . \label{eq_Ei_bis}
\end{align}
Comparisons of (\ref{eq_Ei_bis}) with (\ref{eq_Ei_star}) show that 
$E_i = e_i$ is valid if
$E_{0v} = e_{iv0}-e_{il0}$ and
$E_{0s} = e_{il0}-e_{is0}$, which are indeed the definitions retained in R15.
However, the second line of (\ref{eq_Ei_bis}) must also be neglected.
This is true for $e_{ia0}$, which acts as global constant offset for all species.
Differently, $q_t \: (e_{il0}-e_{ia0})$ can only be neglected for adiabatic (closed) parcels of moist air, namely if $q_a = 1 - q_t$ and $q_t$ are constant with height, or for the assumption $e_{il0} = e_{ia0}$, which is not recalled before Eq.~(1) in R15 and which is not valid.

Therefore, $E_i$ cannot represent the true moist-air internal energy to be used in the general Eq.~(1) of R15, which is called the ``governing equation for internal energy (i.e. the first law of thermodynamics)'' and where the total water content $q_t$ and the diabatic source term $Q$ are {\it a priori\/} different from zero.

The ``equation for enthalpy'' is then defined by Eq.~(2) of R15 in terms of a moist-air specific quantity denoted by $H$ in this comment, leading to
\begin{equation}
 H \; = \; c_{pm} \: (T - T_{\rm trip}) 
   \: + \: q_v \: (E_{0v} + R_v \: T_{\rm trip} )
   \: - \: q_s \: E_{0s}
   \: . \label{eq_H_star}
\end{equation}
This quantity is added to $g \: z$ to form the moist-static energy MSE given by Eqs.~(5)-(6) in R15.

The term $E_{0v} + R_v \: T_{\rm trip} = e_{iv0} + R_v \: T_{\rm trip} - e_{il0}$ 
is equal to $H_{v0}-H_{l0} = L_{\rm vap}(T_{\rm trip})$ in (\ref{eq_H_star}) 
because $e_{iv0} + R_v \: T_{\rm trip} = H_{v0}$ and $e_{il0} = H_{l0}$, 
where the latent heat of vaporization and fusion are 
$L_{\rm vap}(T_{\rm trip}) = H_{v0}-H_{l0}$ and 
$L_{\rm fus}(T_{\rm trip}) = H_{l0}-H_{s0}$, respectively.
Similarly, $E_{0s} = L_{\rm fus}$ because $e_{il0} = H_{l0}$ and $e_{is0} = H_{s0}$.

Let us derive the true moist-air specific enthalpy $h$, to be compared with $H$.
Following the method described in M15 and MG15, the moist-air enthalpy is defined by
\vspace*{-2mm}
\begin{align}
 h & = \;   q_a \: h_{a} 
    \: + \: q_v \: h_{v} 
    \: + \: q_l \: h_{l} 
    \: + \: q_s \: h_{s} 
    \: , \label{eq_H} 
\end{align}
where the partial enthalpies $h_{a}$ to $h_{s}$ can be computed as in (\ref{eq_Eia})-(\ref{eq_Eis}) with heat capacities at constant volume replaced by those at constant pressure.
The main difference with R15 is that reference partial enthalpies $h_{a0}$ to $h_{s0}$ are computed at $T_{\rm trip}$ without further assumptions, leading to
\vspace*{-2mm}
\begin{align}
 h & = \; c_{pm} \: (T - T_{\rm trip}) 
   \: + \: q_v \; L_{\rm vap}(T_{\rm trip})
   \: - \: q_s \; L_{\rm fus}(T_{\rm trip})
  \nonumber \\
  & \quad \: + \: q_t \: (h_{l0}-h_{a0}) \: + \: h_{a0}
   \: . \label{eq_H_bis}
\end{align}

Comparison of (\ref{eq_H_bis}) with (\ref{eq_H_star}) shows that $h = H$ only if the second line in (\ref{eq_H_bis}) is a constant and could be discarded.
Since reference values of thermal enthalpies of dry air $h_{a0}$ and liquid water $h_{l0}$ derived in M15 and MG15 are different from each other, the second line of (\ref{eq_H_star}) can only be discarded for a closed parcel (namely if the specific total-water content $q_t$ is constant with height).

However, the moist-air specific thermal enthalpy $h$ is different from $H$ for an open parcel of fluid, namely for varying values of $q_t$.
In consequence, the quantity $H$ which corresponds to the MSE cannot represent the moist-air specific enthalpy in R15 for all atmospheric conditions and Eq.~(2) in R15 is only valid for a closed parcel of moist air: it does not represent the general governing equation for enthalpy, namely $dh/dt= (\ldots)$.

\vspace*{-2mm}

 \section{The equation for enthalpy} 
\vspace*{-5mm}

It is important to separate the equation for $T$ from the equation for $h$, or the possibility to compute $T$ from the difficulties to compute $h$ itself.

It is recalled in section~2.1 of MG15 that the global offset values $h_{a0}$ in (\ref{eq_H_bis}) must not acquire a physical meaning and that the term $q_t\, (h_{l0}-h_{a0})$ does not need to be computed in the equation for temperature.
However, for open systems and according to \citet{degroot_mazur84}, it is needed to start with the relevant definition (\ref{eq_H_bis}) for $h$ in order to derive relevant versions of the so-called equations $c_p \: dT/dt = (\ldots)$ or $d(c_p \, T)/dt = (\ldots)$.

Indeed, the equation for $T$ can be derived from the one for enthalpy via the cancellation of several terms depending on external changes of dry-air and water contents, and with the appearance of extra terms in the rhs of the equation for $T$.
Therefore, if the term depending on $q_t$ in the second line of (\ref{eq_H_bis}) is missing, it is no longer possible to get the relevant equation for $T$ for open systems.

This issue was already discussed in \citet[ p.158-160]{Richardson22} who imagined some process of adding water-substance reversibly to a given mass of moist-air.
He asked the question: 
{\it what energy (and entropy) are to be ascribed to unit mass of the incoming substance?\/}
Accordingly, the precise computation of $h$ may be useful in order to answer the question: are the enthalpies of two given parcels of moist air different or equal to each other?
As expected, the global offset $h_{a0}$ in (\ref{eq_H_bis}) cancels out and has no physical meaning (just like the arbitrary origin for geopotential).
Differently, the term $q_t \: (h_{l0}-h_{a0})$ gives non-zero impacts if $q_t$ is not the same for the two parcels.

\begin{figure*}[hbt]
\centerline{\includegraphics[width=0.7\linewidth]{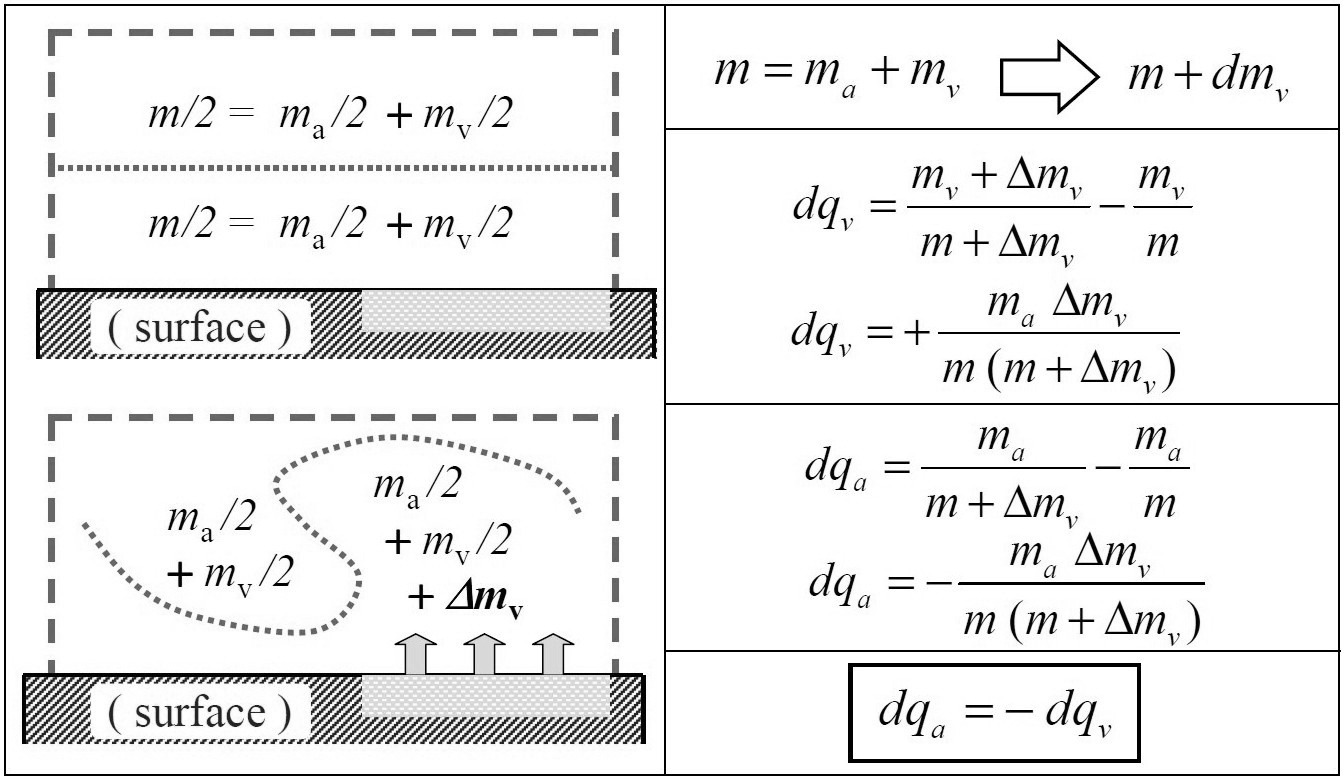}}
\vspace*{-3mm} 
\caption{An explanation for the apparent ``{\it exchange\/}'' of dry air by water vapor during evaporation processes occurring at the boundaries of an open parcel of moist air.
} 
\label{fig1}
\end{figure*}

A way to answer to the question asked by Richardson is illustrated in Fig~\ref{fig1}.
It is shown that the evaporation of a given mass of water $\Delta m_v$ (the incoming substance) inside a given mass $m = m_a + m_v$ of moist air can be interpreted as a replacement of a specific content of dry air $d q_a$ by an opposite specific content of water vapor $dq_v = d q_t = - d q_a$.
The impact on the specific enthalpy is thus equal to $d h = d q_t \: (h_v - h_a)$, which corresponds to the first term in the second line of (\ref{eq_H_bis}) since $(h_v - h_a) = (h_l - h_a) + L_{\rm vap}$.

The evaporation process refers to open-system thermodynamics and there is no attempt to imagine some ``{\it nuclear alchemy\/}'' between dry air and water vapor.
The impact $d h = d q_t \: (h_v - h_a)$ simply corresponds to the opposite (external) changes in specific contents for the two species, changes which may occur at the boundary of the parcel.

The derivation of the moist-air enthalpy given by (\ref{eq_H_bis}) is more direct and avoids the method mentioned in R15, where the Lagrangian derivative of the term 
{$R_m \: T_{\rm trip} - q_v \: R_v \: T_{\rm trip}$}
(indeed equal to zero for adiabatic motions) is added arbitrarily to Eq.~(2) without clear justification: why is this term selected, and not, for example, its double?

Any departure from the adiabatic hypotheses would correspond to varying values of $q_t$ and imply that the second line of (\ref{eq_Ei_bis}) and (\ref{eq_H_bis}) must be taken into account.
This occurs for any realistic core ascents in clouds where some part of the condensed water can be added/withdrawn from the parcel by precipitation.
It is also the observed for the diluted parcels like those studied in \citet{RompsKuang2010}, where the entrainment (or detrainment) processes between the parcel and the environment must lead to varying $q_t$.
However, the second line of (\ref{eq_Ei_bis}) is not considered in $E^{\rm tot}$ in the Appendix of \citet{RompsKuang2010}.

The adiabatic parcel that ascends with the multi-kilometer vertical extent depicted in Fig.2 of R15 must condensate liquid or ice contents of about $20$~g~kg${}^{-1}$ at high levels.
Such large values cannot be reached in realistic clouds, so the test described in R15 must be considered as a pure academic validation of conservative properties.

Since ``open-parcel'' diabatic conditions are always observed in both the atmosphere and the numerical models, it is important for operational purposes to deal with the impact of  precipitation or entrainment/detrainment processes, which cannot be taken into account starting from Eqs.(3), (5) or (6) of R15.

The advantage of keeping all terms in (\ref{eq_H_bis}) and replacing MSE by $h + g \: z$ is that this allows the change in moist-air enthalpy (and then in $h + g \: z$) to be evaluated in all conditions, including those where $q_t$ is varying and where motions are not adiabatic.
In the following section, it is demonstrated that the same method of searching for a general expression for the moist air entropy leads to results which are different from those published in R15, with expected large impact when more realistic pseudo-adiabatic profiles are considered. 

\vspace*{-2mm}

 \section{The moist-air entropy} 
\vspace*{-5mm}

It is explained in section 3 and the appendix of R15 that ``$\theta_e$ is simply the exponential of the (moist-air) entropy,'' although ``$\theta_e$ has been written in many different ways with varying degrees of completeness and accuracy.''
It is shown in this section that the moist-air entropy cannot be written in many different ways and that $\theta_e$ defined in R15 does not represent the general moist-air entropy, due to several arbitrary approximations.

The moist-air entropy corresponding to $\theta_e$ in R15 has previously been computed in \citet{Romps2008} and \citet{RompsKuang2010} starting from Dalton's law
\vspace*{-1mm}
\begin{align}
 s & = \;   q_a \: s_{a} 
    \: + \: q_v \: s_{v} 
    \: + \: q_l \: s_{l} 
    \: + \: q_s \: s_{s} 
    \: \label{eq_S_R15} 
\end{align}
and with partial entropies defined by
\vspace*{-1mm}
\begin{align}
 s_a & = \; 
      c_{pa} \: \log(T/T_{\rm trip}) 
    - R_a    \: \log(p_a/p_{\rm trip})
    \: + \: s_{0a}
    \: , \label{eq_sa} \\
 s_v & = \; 
      c_{pv} \: \log(T/T_{\rm trip}) 
    - R_a    \: \log(p_v/p_{\rm trip})
    \: + \: s_{0v}
    \: , \label{eq_sv} \\
 s_l & = \; 
      c_{pl} \: \log(T/T_{\rm trip})
    \: + \: s_{0l}
    \: , \label{eq_sl} \\
 s_s & = \; 
      c_{ps} \: \log(T/T_{\rm trip}) 
    \: - \: s_{0s}
    \: , \label{eq_ss}
\end{align}
where the triple-point conditions are $T_{\rm trip} = 273.16$~K and  $p_{\rm trip} = 6.12$~hPa.
It is arbitrarily assumed in \citet{Romps2008} that
$s_{0a} = 0$,
$s_{0l} = 0$,
$s_{0v} = R_v + E_{0v}/T_{\rm trip} = L_{\rm vap}(T_{\rm trip}) / T_{\rm trip}$ and
$s_{0s} = E_{0s}/T_{\rm trip}  = L_{\rm fus}(T_{\rm trip}) / T_{\rm trip}$.

These assumptions are similar to those made in \citet{Emanuel94} and \citet{Pauluis_al_2010}, but they all contradict the third law of thermodynamics, which states that the entropy of any substance is equal to a universal constant (set to zero) for the most stable crystalline form of the substance and at absolute zero temperature (namely for $T=0$~K different from $T_{\rm trip}$ and independently for all substances).
More on this important issue will be discussed in the conclusion.

By making these arbitrary choices, the potential temperature $\theta_e$ is then derived in \citet{RompsKuang2010} from (\ref{eq_S_R15})-(\ref{eq_ss}) by writing
\vspace*{-2mm}
\begin{align}
    (s)_{\rm R15}
    & = \; s_{ref} \: + \: 
    q_a \; c_{pa} \; \ln(\theta_e)
    \: , \label{eq_S_thetae_R15} 
\end{align}
where the reference value is arbitrarily set to 
\vspace*{-2mm}
\begin{align}
    s_{ref}
    & = \; q_a \; c_{pa} \; 
        \ln\left[ 
               \; T_{\rm trip} 
            \: \left(
                  {p_0}/{p_{\rm trip}}
            \right)^{R_a/c_{pa}} 
        \: \right]
    \: \label{eq_Sref_thetae_R15} 
\end{align}
and with $p_0 = 1000$~hPa.
This choice for $s_{ref}$ is not justified in R15, and appears to be motivated by the desire to arrive at a certain result in Eq.~(A1), which can be rewritten as
\vspace*{-2mm}
\begin{align}
  {\theta}_e &  \: = \:  
   \theta \;
   \exp \!\left(
     \:   
     \frac{r_v \; L_{\rm vap}(T_{\rm trip})  - r_s \; L_{\rm fus}(T_{\rm trip})}{{c}_{pa} \: T_{\rm trip}}
   \: \right)
 \nonumber \\
    &  \quad \times
     \left( \frac{p}{p_a}\right)^{\!R_a/{c}_{pa}}
     \left( \frac{p_{\rm trip}}{p_v}\right)^{\!r_v\:R_v/{c}_{pa}}
 \nonumber \\
    &  \quad \times
     \left( \frac{T}{T_{\rm trip}}
     \right)^{\!(r_v\:{c}_{pv}+r_l\:{c}_{pl}+r_s\:{c}_{ps})/{c}_{pa}}
     \! \! 
  \: .
 \label{eq_theta_e_R15}
\end{align}
The dry-air potential temperature $\theta =  T \: (p_0/p)^{R_a/C_{pa}}$ is not explicitly included in Eq.~(A1) of R15, which is however equivalent to (\ref{eq_theta_e_R15}) due to the extra term $(p/p_a)^{R_a/{c}_{pa}}$.
Moreover, the terms $s_{0v}$ and $s_{0s}$ in R15 are replaced in (\ref{eq_theta_e_R15}) by the latent heat of vaporization and fusion computed at $T_{\rm trip}$ and divided by $T_{\rm trip}$.
The alternative formulation 
\vspace*{-2mm}
\begin{align}
  {\theta}_e &  \: = \:  
   \theta \;
   \exp \!\left(
     \: - \:
     \frac{r_l \; L_{\rm vap}(T_r) + r_s \; L_{\rm sub}(T_r)}{{c}_{pa} \: T_r}
   \: \right)
 \nonumber \\
    &  \quad \times \;
   \exp \!\left(
     \:
     \frac{ L_{\rm vap}(T_r) }{{c}_{pa} \: T_r} \: r_t
   \: \right)
 \nonumber \\
    &  \quad \times
     \left( \frac{T}{T_r}
     \right)^{\!(r_v\:{c}_{pv}+r_l\:{c}_{pl}+r_s\:{c}_{ps})/{c}_{pa}} \:
     \left( \frac{p_r}{p}\right)^{\!\gamma \, r_v}
 \nonumber \\
    &  \quad \times
           \:  \frac{(1\!+\!\eta\,r_v)^{\kappa \,+ \gamma \:r_v}}
                    {(\eta\,r_v)^{\,\gamma \:r_v}}
           \;
           \:  \frac{(\eta\,r_r)^{\,\gamma \:r_v}}
                    {(1\!+\!\eta\,r_r)^{\gamma \:r_v}}
  \:
 \label{eq_theta_e_R15_bis}
\end{align}
is written in such a way as to be more easily compared with other published formulations. 
It is obtained by using
$T_r = T_{\rm trip}$,
$e_r = p_{\rm trip}$,
$p_r = p_0$,
$p/p_a = 1+\eta\: r_v$.
Moreover
${p_{\rm trip}}/{p_v} = (e_r/p_r) \: (p_r/p) \: (p/p_v)$
is computed from 
$e_r/p_r = (\eta \: r_r)/(1+\eta\: r_r)$,
$r_r = \varepsilon \: e_r \, / \,  (p_r - e_r)$ and
$p/p_v = (1+\eta\: r_v)/(\eta \: r_v)$,
with $\gamma = R_v/{c}_{pa} \approx 0.46$,
$\eta = R_v/R_a \approx 1.608$,
and $\varepsilon = R_a/R_v \approx 0.622$.
The term $r_t \: L_{\rm vap}(T_r)$ is added to form 
the second exponential term and subtracted from the first exponential term, 
with the corresponding change of
$r_v \; L_{\rm vap}(T_r) - r_s \; L_{\rm fus}(T_r)$
into 
$-r_l \; L_{\rm vap}(T_r) - r_s \; L_{\rm sub}(T_r)$.

It is explained in \citet[hereafter M11]{Marquet11} that it is possible to compute the moist-air entropy without making the assumptions needed to arrive at (\ref{eq_theta_e_R15})-(\ref{eq_theta_e_R15_bis}).
The method is to start with the same Dalton's law as (\ref{eq_S_R15}), but with the partial entropies written as
\vspace*{-4mm}
\begin{align}
 s_a & = \; 
      c_{pa} \: \log(T/T_r) 
    - R_a    \: \log(p_a/p_{ar})
    \: + \: s_{ar}
    \: , \label{eq_sar} \\
 s_v & = \; 
      c_{pv} \: \log(T/T_r) 
    - R_a    \: \log(p_v/p_{vr})
    \: + \: s_{vr}
    \: , \label{eq_svr} \\
 s_l & = \; 
      c_{pl} \: \log(T/T_r)
    \: + \: s_{lr}
    \: , \label{eq_slr} \\
 s_s & = \; 
      c_{ps} \: \log(T/T_r) 
    \: + \: s_{sr}
    \: . \label{eq_ssr}
\end{align}
The reference entropies  $s_{ar}(T_r,p_{ar})$, $s_{vr}(T_r,p_{vr})$, $s_{lr}(T_r)$ and $s_{sr}(T_r)$ are not set to prescribed values and are thus different from those in (\ref{eq_sa})-(\ref{eq_ss}).

The moist-air entropy can then be expressed in terms of a general potential temperature $\theta_s$, leading to
\vspace*{-4mm}
\begin{align}
  (s)_{\rm M11} &  \: = \: s_{ref}  \; + \; c_{pa} \: \ln(\theta_{s}) 
  \: ,
 \label{eq_s_theta_s} \\
  {\theta}_{s} &  \: = \:  
   \theta \;
   \exp \!\left(
     \:  - \:   
     \frac{ q_l \; L_{\rm vap}(T) + q_s \; L_{\rm sub}(T)}{{c}_{pa}\:T}
   \: \right)
 \nonumber \\
    &  \quad \times \;
   \: \exp\! \left(  \Lambda_r \: q_t  \right) \; \;
     \left( \frac{T}{T_r}\right)^{\!\lambda \,q_t} \; \;
     \! \! \left( \frac{p_r}{p}\right)^{\!(\gamma - \kappa) \,q_t}
 \nonumber \\
    &  \quad \times
           \:  \frac{(1\!+\!\eta\,r_v)^{\,\kappa + \, (\gamma - \kappa) \, q_t}}
                    {(\eta\,r_v)^{\gamma\,q_t}}
           \;
           \:  \frac{(\eta\,r_r)^{\gamma\,q_t}}
                    {(1\!+\!\eta\,r_r)^{(\gamma - \kappa) \, q_t}}
  \: .
 \label{eq_theta_s}
\end{align}
In contrast with (\ref{eq_S_thetae_R15}) in R15, it is shown in M11 that the terms 
$s_{ref} = 1139$~J~K${}^{-1}$~kg${}^{-1}$ 
and
$c_{pa} \approx 1005$~J~K${}^{-1}$~kg${}^{-1}$ 
appearing in (\ref{eq_s_theta_s}) 
are two constants.
This justifies the use of $\theta_s$, given by (\ref{eq_theta_s}), as a true equivalent of the moist-air entropy regardless of the atmospheric conditions, in particular with or without the adiabatic assumption and including the case of varying values of $q_t = 1 - q_a$.

This is a clear advantage with respect to the formulation published in \citet{HaufHoller87},  \citet{Marquet93},  \citet{Emanuel94} or R15, where a portion of moisture variables $q_t$ are located outside the logarithm, thus preventing the previous moist-air potential temperature from being truly equivalent to the moist-air entropy, including $\theta_e$ given by (\ref{eq_theta_e_R15})-(\ref{eq_theta_e_R15_bis}).

The moist-air entropy potential temperature $\theta_s$ depends on the absolute temperature $T$, the total pressure $p$, the total-water specific content $q_t = q_v + q_l + q_s$ and the water vapor mixing ratio $r_v$.
The thermodynamic constants are the same as in (\ref{eq_theta_e_R15_bis}), plus
$\kappa = R_a/{c}_{pa} \approx 0.286$ and
$\lambda = {c}_{pv}/{c}_{pa} - 1 \approx 0.838$.

The reference temperature and total pressure are set to $T_r=273.15$~K and $p_r = 1000$~hPa in M11.
The reference partial pressure $e_r = 6.11$~hPa is the saturating pressure at $T_r$ and $p_r$.
It is shown in M11 that $s_{ref}$ is indeed a constant and $\theta_s$ is independent of the choice of the reference values $T_r$ and $p_r$ if the reference mixing ratio is logically defined by $r_r(T_r,p_r) = \varepsilon \: e_r(T_r) \, / \,  [\: p_r - e_r(T_r) \: ] \approx 3.82 $~g~kg${}^{-1}$.

The new term
$\Lambda_r = (  \: s_{vr} - s_{ar} ) \, / \, c_{pa}$
depends on the reference entropies of dry air and water vapor
$s_{vr}(T_r, e_r) \approx 12673 $~J~K${}^{-1}$ and 
$s_{ar}(T_r, p_r-e_r) \approx  6777$~J~K${}^{-1}$,
which correspond to values published in \citet{HaufHoller87} and M11 and determined from usual thermodynamic tables, leading to the value $\Lambda_r \approx 5.87$.
The same reference values for entropies are explicitly computed in M15 from the third law of thermodynamics and by using accurate cryogenic datasets for N$_2$, O$_2$ and H$_2$O.

It is now possible to compare $\theta_e$ given by (\ref{eq_theta_e_R15_bis}) and $\theta_s$ given by (\ref{eq_theta_s}).
The differences are:
\begin{itemize}[label=$\bullet$,leftmargin=3mm,parsep=0.0cm,itemsep=0.0cm,topsep=0.02cm,rightmargin=2mm]
   \item 
mixing ratios in the exponential terms in (\ref{eq_theta_e_R15_bis}) are replaced by specific contents in (\ref{eq_theta_s})
   \item 
latent heat is computed at $T_r$ in (\ref{eq_theta_e_R15_bis}) and at $T$ in (\ref{eq_theta_s});
   \item 
the factor $L_{\rm vap}(T_r)/({c}_{pa}\:T_r) \approx 9$ in the second exponential in (\ref{eq_theta_e_R15_bis}) is replaced by $\Lambda_r \approx 6$ in (\ref{eq_theta_s});
   \item 
exponents in other terms depend on the mixing ratios ($r_v, r_l, r_s$) in (\ref{eq_theta_e_R15_bis}), whereas they all depend on $q_t$ in (\ref{eq_theta_s}), with different thermodynamic constants.
This means that the terms depending on $(T/T_r)$, $(p_r/p)$, $(1+\eta\:r_v)$ and $(\eta\: r_v)$, and in particular those depending on $p(z)$ and $r_v(z)$, vary according to height differently in (\ref{eq_theta_e_R15_bis}) and in (\ref{eq_theta_s}).
\end{itemize}

It is posible to compare the entropies $(s)_{\rm M11}$ and $(s)_{\rm R15}$ themselves, since they can be expressed by the exact and simple relation
\vspace*{-2mm}
\begin{align}
    (s)_{\rm M11}  & \: = \; (s)_{\rm R15}
     \: + \: s_1 \: + \: s_2 
    \: , \label{eq_S_M11_R15} \\
     s_1  & \: = \; - \: q_t \: (s_{dr}-s_{lr})
    \: , \label{eq_S_M11_R15_S1} \\
     s_2  & \: = \; 
       (1-q_t) \: \ln(p_{ar}/p_{vr})
    \: , \label{eq_S_M11_R15_S2} 
\end{align}
where the constant reference values are
$T_r = 273.16$~K, $p_{r}  = 1000$~hPa,
$p_{vr} = p_{\rm trip} \approx 6.11$~hPa,
$p_{ar} = p_r - p_{\rm trip} \approx 994$~hPa,
$s_{dr} = 6777$~J~K${}^{-1}$~kg${}^{-1}$ and
$s_{lr} = 3518$~J~K${}^{-1}$~kg${}^{-1}$,

Since $(s_{dr}-s_{lr})$ and $\ln(p_{ar}/p_{vr})$ are constant, $(s)_{\rm M11}$ and $(s)_{\rm R15}$ are equivalent up to the constant sum $s_1 \: + \: s_2$ only if $q_t =  1 - q_a$ is constant with height, namely for closed parcels of moist-air.
However, if $q_t =  1 - q_a$ varies with time and/or with space, the difference $(s)_{\rm M11} \: - \; (s)_{\rm R15}$ is equal to the sum $s_1 \: + \: s_2$ which varies with time and/or with space.
This means that $(s)_{\rm R15}$ is not a  measure of the entropy for open parcels of moist-air.

 \section{The moist-air adiabatic profile} 
\vspace*{-5mm}

\begin{table*}
\caption{\small \it Thermodynamic conditions of the saturated adiabatic updraft starting at $z=0$~m and $p=1000$~hPa with a temperature of $300.5$~K:
height $z$ in m, pressure $p$ in hPa, temperature $T$ in K,
specific contents $q_v$, $q_l$ and $q_s$ in g~kg${}^{-1}$,
potential temperatures $\theta_s$, $\theta_e$
and $\mbox{MSE}/R_d$ in K.
\label{Table}
}
\vspace*{-0mm} 
\centering
\vspace*{2mm}
\begin{tabular}{||c||c|c||c|c||}
\hline
   $z$  &    $0$    &  $6593.8$ & $6594.6$  &  $16980$  \\ 
\hline \hline
   $p$  &  $1000$   & $456.95$  & $456.90$  &   $100$   \\ 
\hline
   $T$  &  $300.5$  & $273.162$ & $273.158$ &  $196.9$  \\ 
\hline
  $q_v$ &  $22.902$ & $8.240$   &  $8.238$  &  $0.006$  \\
\hline
  $q_l$ &    $0$    &  $14.663$  &   $0$     &    $0$    \\
\hline
  $q_s$ &    $0$    &    $0$    & $14.665$   &  $22.896$   \\
\hline
 $(\theta_s)_{M11}$ 
        & $341.482$ & $341.482$ & $335.446$ & $335.446$ \\ 
\hline
 $(\theta_s)_{M11}$ 
        &           &          & $(341.482)$ & $(341.481)$ \\ 
\hline
 $(\theta_e)_{R15}$ 
        & $370.376$ & $370.376$  & $363.678$  & $363.678$ \\ 
\hline
 $(\theta_e)_{R15}$ 
        &           &          & $(370.376)$ & $(370.375)$ \\ 
\hline
 $(\mbox{MSE}/R_d)_{R15}$ 
        & $297.046$ & $297.048$ & $279.999$ & $280.019$ \\ 
\hline
 $(\mbox{MSE}/R_d)_{R15}$ 
        &           &          & $(297.048)$ & $(297.066)$ \\ 
\hline
\end{tabular}
\vspace*{-4mm} 
\end{table*}

The impact of approximations made in R15 can be studied by building the same saturated moist-air adiabatic vertical profile described in R15 starting at $z=0$~m and $p=1000$~hPa with an initial temperature of $300.5$~K.

Since the aim of this section is to compare $\theta_s$ and $\theta_e$ for a parcel undergoing isentropic transformations, it is important to use a definition of the moist-air entropy which is independent of the choices of $\theta_s$ and $\theta_e$.

The choice retained in R15 for defining the moist-air entropy is not explicitly described.
It is likely based on the formulas (\ref{eq_S_thetae_R15}) and (\ref{eq_Sref_thetae_R15}) where $q_a = 1-q_t$ and $s_{ref}$ are constant with height, leading to a moist-air entropy defined by $q_a \: c_{pd} \ln (\theta_e)$ up to a constant term and where $\theta_e$ is given by (\ref{eq_theta_e_R15}).

Differently, the saturated adiabatic lapse rate retained in this section is defined by the exact differential Eqs.~(3) and (4) given in \citet{Saunders57}.
It can be shown that these equations corresponds to Eq.~(4) in \citet{Geleyn_Marquet12}, which corresponds exactly to Eq.~(16) in \citet{Marquet_Geleyn13} and to
\vspace*{-4mm}
\begin{align}
 \Gamma_{\rm adiab}  & 
 \: = \; - \: 
  \left. \frac{\partial T}{\partial z} \right|_{(s, \, q_t)}
  = \;  
   \frac{g}{c_{p}}
     \left(
     \frac
      { 
      1 + \displaystyle{
             \frac{L_x \: r_{x}}
                  {R_a \: T}
                       }
      }
      {
      1 + \displaystyle{
             \frac{R \; L_x^2 \:r_{x}}
                  {R_a \: c_{p} \: R_v \: T^2}
                       }
      }
     \right)
  .
 \label{eq_grad_adiab}
\end{align}
$L_x$ and $r_{x}$ are notations for the latent heat $L_{\rm vap}$ and the saturating mixing ratio (over liquid water) $r_{vl}$  for $T \geq T_{\rm trip}$, or for $L_{\rm sub}$ and $r_{vs}$ (over ice) for $T < T_{\rm trip}$.
In contrast, the lapse rates computed in \citet{DK82} and \citet{Emanuel94} are not computed with the relevant moist-air entropy and they disagree with Eqs.~(3) and (4) in \citet{Saunders57}.

\begin{figure}[hbt]
 \centerline{\includegraphics[width=0.99\linewidth]{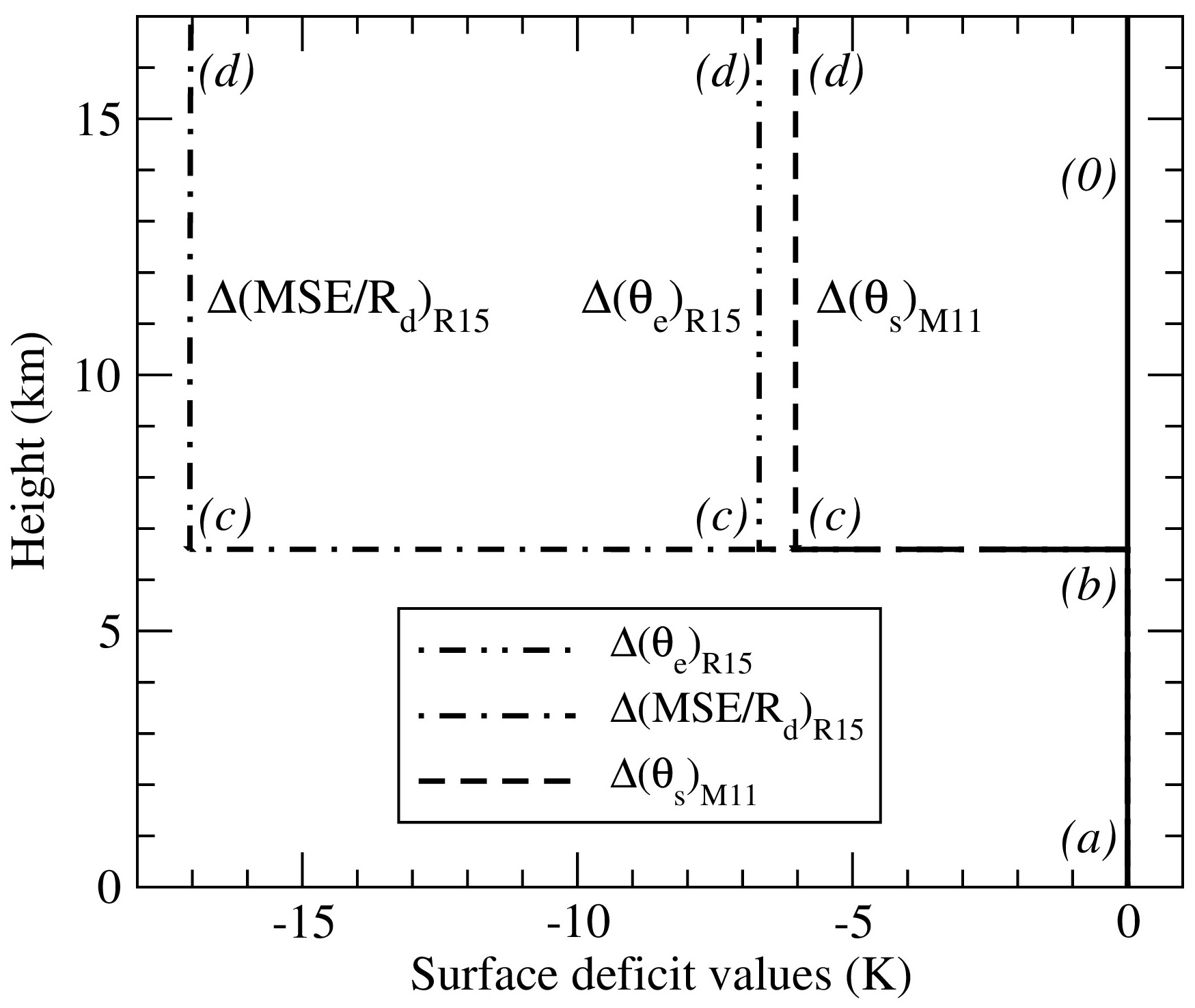}}
\vspace*{-3mm} 
\caption{\small \it 
The surface deficit values in $(\theta_e)_{\rm R15}$ (double-dotted dashed), $(\mbox{MSE}/R_d)_{\rm R15}$ (dotted-dashed) and $(\theta_s)_{\rm M11}$ (dashed) plotted in terms of the height (in km) above the surface level and for the adiabatic vertical profile.
Units of all surface deficit values are in K.
The vertical lines are discontinuous at the triple-point temperature level at $6596$~m.
} 
\label{fig2}
\end{figure}

The moist-adiabatic (isentropic) vertical ascent is computed by integrating (\ref{eq_grad_adiab}) with an interval of $0.05$~hPa between $1000$ and $100$~hPa (use of an accurate leap-frog scheme with an Asselin's filter).
Results are shown in Table~\ref{Table} and Fig~\ref{fig2}.
Here, in contrast to R15, but in agreement with Fig~8 in \citet{RompsKuang2010}, at the triple-point temperature the liquid water is suddenly frozen.

The main result is the expected adiabatic conservative feature observed in Fig~\ref{fig2} for \mbox{MSE} or $h+\phi$ (dotted-dashed line), $\theta_s$ (dashed line) and $\theta_e$ (double-dotted dashed line) for each domain $T<T_{\rm trip}$ and $T>T_{\rm trip}$.
This can be explained by the adiabatic relationship recalled in \citet{Ambaum10} and M15: $0 = \partial s/\partial z = (c_{pd}/\theta_s) \: \partial\theta_s/\partial z \approx (1/T) \: \partial /\partial z (h + \phi)$, where ``$h+\phi$'' is the generalized enthalpy and where the specific enthalpy $h$ is given by (\ref{eq_H_bis}).

Since $\theta_s$, $\theta_e$ and $q_a =  1 - q_t$ are constant with height above and below the freezing level, the two entropies  $(s)_{\rm R15}(\theta_e)$ and $(s)_{\rm M11}(\theta_s)$ given by (\ref{eq_S_thetae_R15}) and (\ref{eq_s_theta_s}) and linked by (\ref{eq_S_M11_R15})-(\ref{eq_S_M11_R15_S2}) are also constant with height above and below the freezing level for the adiabatic profile in Fig~\ref{fig2}.

\begin{figure*}[hbt]
 \centerline{\includegraphics[width=0.65\linewidth]{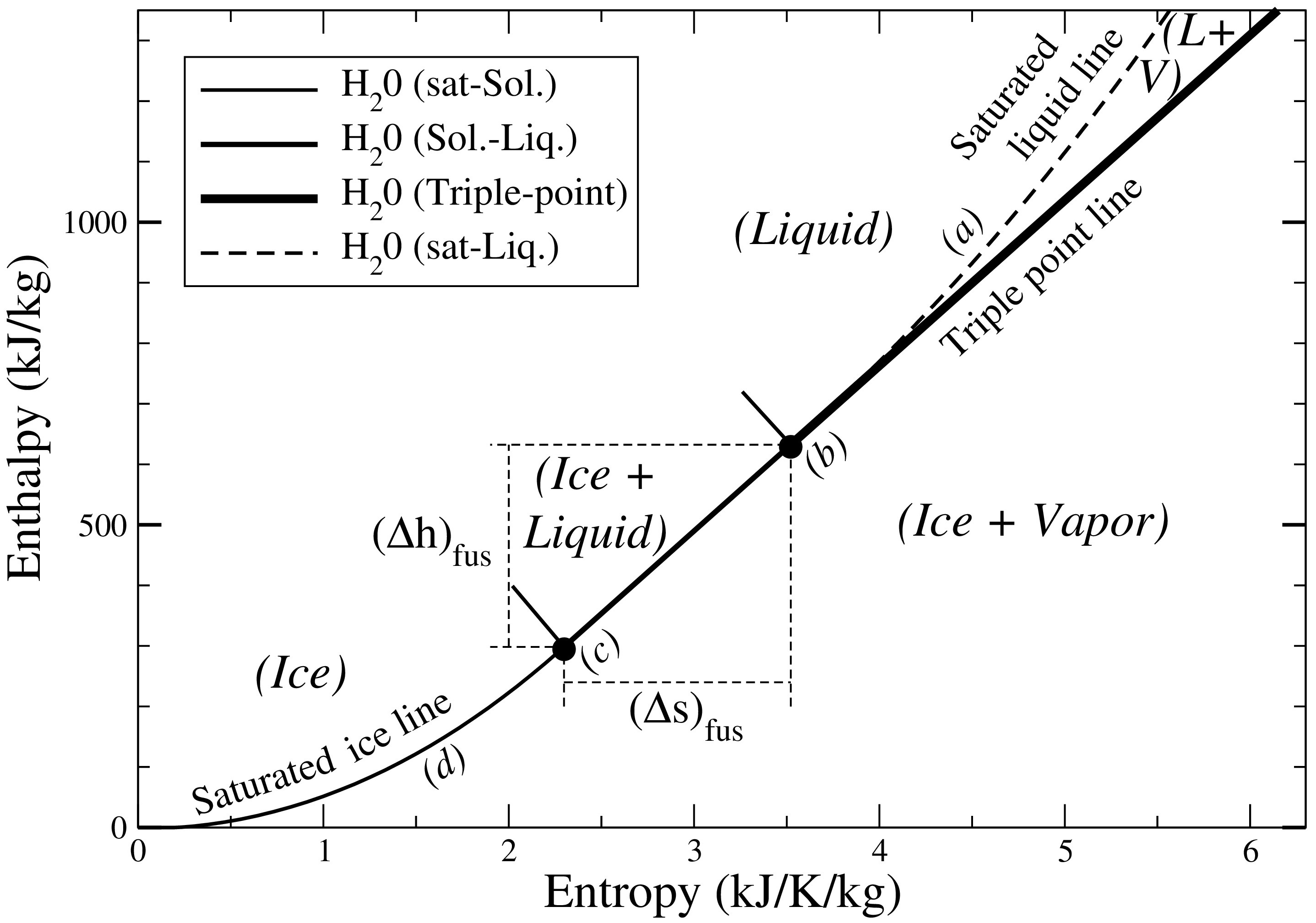}}
\vspace*{-3mm} 
\caption{\small \it 
The Mollier (or specific ``enthalpy-entropy'') diagram (or chart) for the three water species: ice (Ih), liquid and vapor.
The specific entropy and enthalpy are plotted from $0$~K to $450$~K.
The liquid and vapor (L$+$V) domain is located in between the ``saturated liquid''  and ``Triple point'' lines (upper right corner).
} 
\label{fig3}
\end{figure*}

However, the values below the freezing level are not continuous with those above this level, where liquid water is suddenly frozen.
Similar discontinuous features are shown in Fig.~8 in \citet{RompsKuang2010} at about $4$~km for the parcel buoyancy $b(z)$ and in Fig.~2 (right) in R15 at about $6$~km for $\Delta(T_v)$.
These jumps in $b(z)$ and $\Delta(T_v)$ are relevant.
They corresponds to the impact of the solidification of existing cloud liquid water at these levels.

These discontinuities are smoothed in Fig.~2 (right) R15 and in Figs.~11 of \citet{RompsKuang2010} by imposing a linear transition between liquid water and ice, in order to mimic observations of supercooled water and of a mixed-phase in deep convective cloud.
However, the smoothing is not complete in R15, since a hook is still observed in Fig.~2 (right) within the isothermal layer close to the freezing level at about $6.6$~km.
Another hook is observed at about $12.5$~km, at the top of the mixed-phase at the temperature of $240$~K.

In order to better understand the physical meaning of these discontinuities of hooks (namely the jumps in both enthalpy and entropy), it is useful to plot in Fig~\ref{fig3} the enthalpy-entropy chart for water
\citep{Mollier27,Bejan88}.
The curve for ice (Ih) between $0$~K and the triple point temperature is plotted with values of $s(T)$ and $h(T)$  computed in \citet{Marquet11,Marquet15a}.


The saturated adiabatic vertical profile considered in Fig.~\ref{fig2} corresponds to the continuous path: 
(a) $\rightarrow$ (b) $\rightarrow$ (c) $\rightarrow$ (d).
The discontinuous and negative jumps in MSE, $\theta_e$ or $\theta_s$ observed at the triple point temperature in Fig.~\ref{fig2} correspond to the continuous and negative changes in entropy $(\Delta s)_{\rm sol} = - \: (\Delta s)_{\rm fus}$ 
and enthalpy $(\Delta h)_{\rm sol} = - \: (\Delta h)_{\rm fus}$ in the Mollier chart.
These changes are both associated with the continuous step (b) $\rightarrow$ (c) in Fig.~\ref{fig3}, which represents the impact of the solidification of liquid water into ice at the constant triple point temperature.
It is a straight line with a constant slope of $T_{\rm trip}$ because $(\Delta h)_{\rm fus} = T_{\rm trip} \:  (\Delta s)_{\rm fus}$.

During the two steps (a) $\rightarrow$ (b) and (c) $\rightarrow$ (d) liquid water and ice are in equilibrium with the saturation vapor.
During these steps, since the temperature $T$ varies continuously with height, the enthalpy $h(z)$ and entropy $s(z)$ are continuous functions of $z$.
Differently, the apparent discontinuous jumps in MSE$\: (z)$, $\theta_e(z)$ and $\theta_s(z)$ are explained by the temperature $T_{\rm trip}$ which remains constant during the solidification step (b) $\rightarrow$ (c), which  occurs at the freezing level close to $6.6$~km.

It is however possible, if needed, to add correction terms to remove these discontinuities.
The impact of the irreversible freezing of the content $q_{l0} = 14.664$~g~kg${}^{-1}$ of liquid water at $T_{\rm trip}=273.16$~K corresponds to an increase in enthalpy of $\Delta H = L_{\rm fus} \: \times \: q_{l0} = 4893.377$~J~kg${}^{-1}$, to be added to MSE above the freezing level.

This increase in enthalpy corresponds to changes in potential temperatures and, according to (\ref{eq_s_theta_s}) and (\ref{eq_S_thetae_R15}), $\theta_s$ and $\theta_e$ must be multiplied above the freezing level by the factors $F_s = \exp[ \: \Delta H/ (c_{pd}\:T_{\rm trip}) \:] = 1.0179899$ and $F_e = \exp[ \: \Delta H/ (q_a \: c_{pd}\: T_{\rm trip}) \:] = 1.0184154$, respectively.

If these correction terms are taken into account (see values in parentheses in Table~\ref{Table}), the results $\Delta(\mbox{MSE}/R_d) \approx 0$, $\Delta(\theta_e) \approx 0$ and $\Delta(\theta_s) \approx 0$ are valid, with good accuracy (better than $0.001$~K for the potential temperatures) from the surface up to $17$~km. 
The numerical round-off error is higher for MSE, due to the accumulated errors in $\phi = g \: z$ at high levels.

This proves that any of MSE, $\theta_e$ or $\theta_s$ can be used to built accurate moist-air adiabatic profiles, including the impact of freezing of liquid water species if needed, if the latent heat release can be taken into account at each level where solidification occur, via correction terms like $\Delta H$, $F_e$ and $F_s$.

However, the difference between $F_e$ and $F_s$ depends on $q_a = 1 - q_t$, and thus on the local thermodynamics conditions.
Therefore, the way the potential temperatures are defined (the choice of either $\theta_e$ or $\theta_s$, for instance) may modify the physical meaning of adiabatic vertical profiles.
This cannot be true, and it is clearly shown in next section that only $\theta_s$ is a true measure of the moist-air entropy.



 \section{The moist-air pseudo-adiabatic profile} 
\vspace*{-5mm}

A saturated adiabatic ascent up to $17$~km generates unrealistic (too large) liquid water or ice content in clouds.
Real atmospheric profiles are much closer to pseudo-adiabatic conditions, where the precipitations are completely withdrawn from the updraft.
Accordingly, behavior halfway between adiabatic and pseudo adiabatic conditions with entrainment rates are studied in \citet{RompsKuang2010} .

Moreover, it is suggested in the conclusion of R15 that the same result (namely the conservation of ``MSE$\: + \:$CAPE'') must hold for entraining parcels or parcels that lose condensates by fallout.
This means that the pair-wise comparisons made in R15 between vertical profiles of $\theta_e$, MSE or ``MSE$\: + \:$CAPE'' might be redone for pseudo-adiabatic conditions.
It is thus important to plot and compare previous values of 
$(\mbox{MSE})_{\rm R15}$, $(\mbox{MSE})_{\rm M15}$, $(\theta_e)_{\rm R15}$ and $(\theta_s)_{\rm M11}$
for a moist-air pseudo-adiabatic vertical profile, together with the moist-air entropies $(s)_{\rm R15}$, $s_1$,  $s_2$ and $(s)_{\rm M11}$.

The pseudo-adiabatic vertical profile starts at $z=0$~m and $p=1000$~hPa with the same initial temperature of $300.5$~K as for the adiabatic profile and with RH$\: = 1$.
The pseudo-adiabatic lapse rate defined by Eqs.~(1) and (2) in \citet{Saunders57} corresponds to
\vspace*{-2mm}
\begin{align}
 \Gamma_{\rm pseudo}  & 
 \: = \; 
   \frac{g}{c_{p}}
     \left(
     \frac
      { 
      1 + \displaystyle{
             \frac{L_x \: r_{x}}
                  {R_a \: T}
                       }
      }
      {
      1 - \displaystyle{
             \frac{r_x \: c_x}
                    {c_{p}}
                             }
         + \displaystyle{
             \frac{R \; L_x^2 \:r_{x}}
                  {R_a \: c_{p} \: R_v \: T^2}
                       }
      }
     \right)
  ,
 \label{eq_grad_pseudo_adiab}
\end{align}
where $c_x$ and $r_{x}$ are notations for $c_{pl}$ and $r_{vl}$ above the triple point,  or for $ c_{ps}$ if $r_{vs}$ below the triple point.
The negative extra term in the denominator of (\ref{eq_grad_pseudo_adiab}) explains that $\Gamma_{\rm pseudo} > \Gamma_{\rm adiab}$, leading to colder temperatures in pseudo-adiabatic ascent than for pure adiabatic vertical gradients.

\begin{figure}[hbt]
 \centerline{\includegraphics[width=0.99\linewidth]{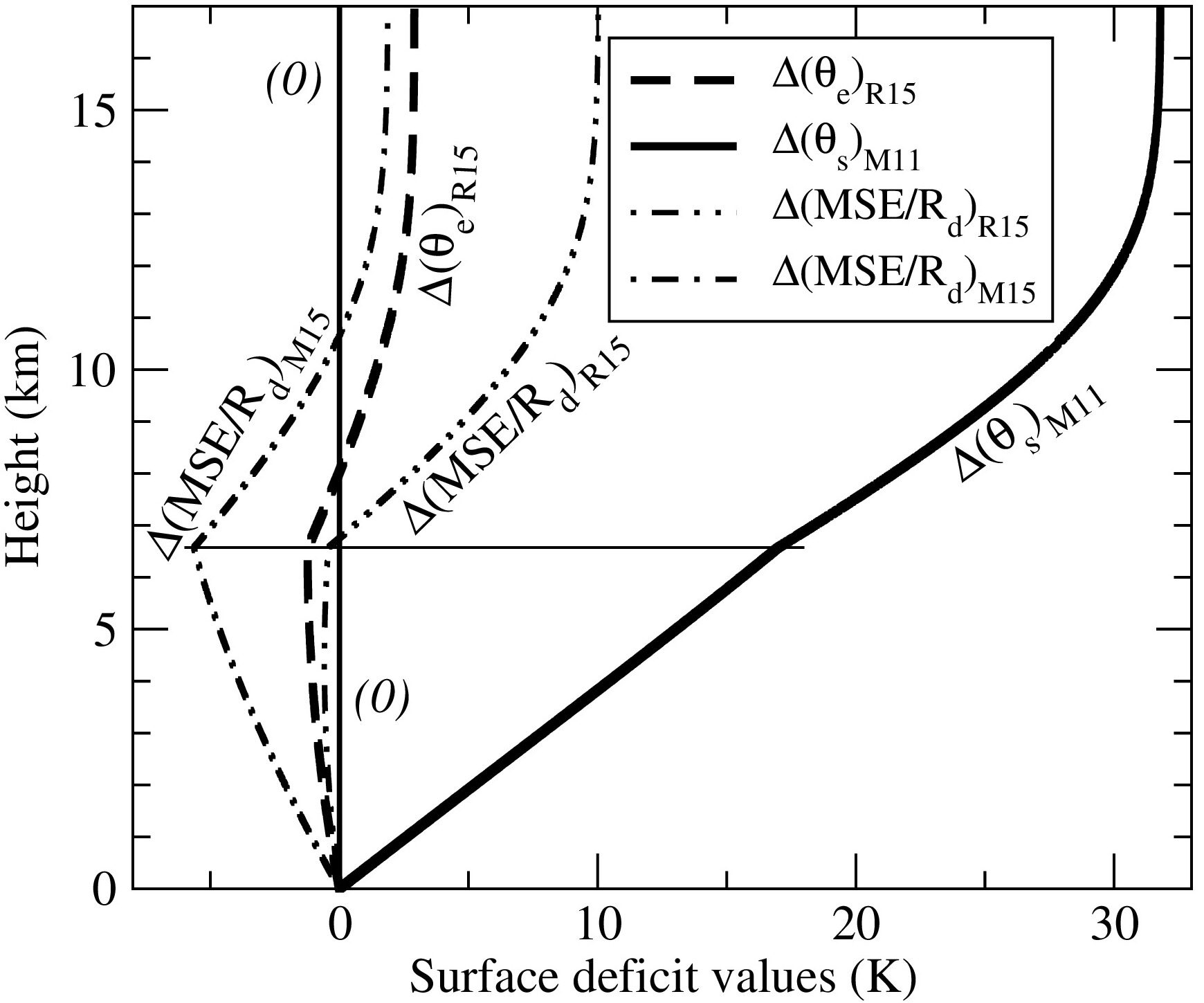}}
\vspace*{-3mm} 
\caption{\small \it 
Same as in Fig.~\ref{fig2} but for the pseudo-adiabatic vertical profile and for the surface deficit values in $(\theta_e)_{\rm R15}$ (double-dotted dashed), $(\mbox{MSE}/R_d)_{\rm R15}$ (dotted-dashed), $(\theta_s)_{\rm M11}$ (dashed) and $(\mbox{MSE}/R_d)_{\rm M15}$ (solid).
The horizontal line denotes the triple-point temperature level at $6571$~m.
} 
\label{fig4}
\end{figure}

The surface deficit values of MSEs and potential temperatures are plotted in Fig~\ref{fig4}.
Logically, none of these quantities are conserved for the pseudo-adiabatic processes.
Differences are clearly observed between $\Delta(\mbox{MSE})_{\rm R15}$ and $\Delta(\mbox{MSE})_{\rm M15}$, due to the impact of the second line in (\ref{eq_H_bis}) and since the saturation water vapor content $q_v =  q_t$ decreases with height for pseudo-diabatic processes.
This means that the way MSE is defined may impact the conserved quantity ``MSE$\: + \:$CAPE'' considered in R15 for open-system processes (namely for entraining parcels or parcels that lose condensates by fallout).

Larger differences are observed between the values of $\Delta(\theta_s)_{\rm M11}$ which increases with height up to $17$~km and those of $\Delta(\theta_e)_{\rm R15}$ which are first decreasing with height below the freezing level at $6.6$~km, and then slightly increases above this level.
This means that at least one of the potential temperatures $\theta_e$ or $\theta_s$ is not valid for describing pseudo-adiabatic processes.

\begin{figure}[hbt]
 \centerline{\includegraphics[width=0.99\linewidth]{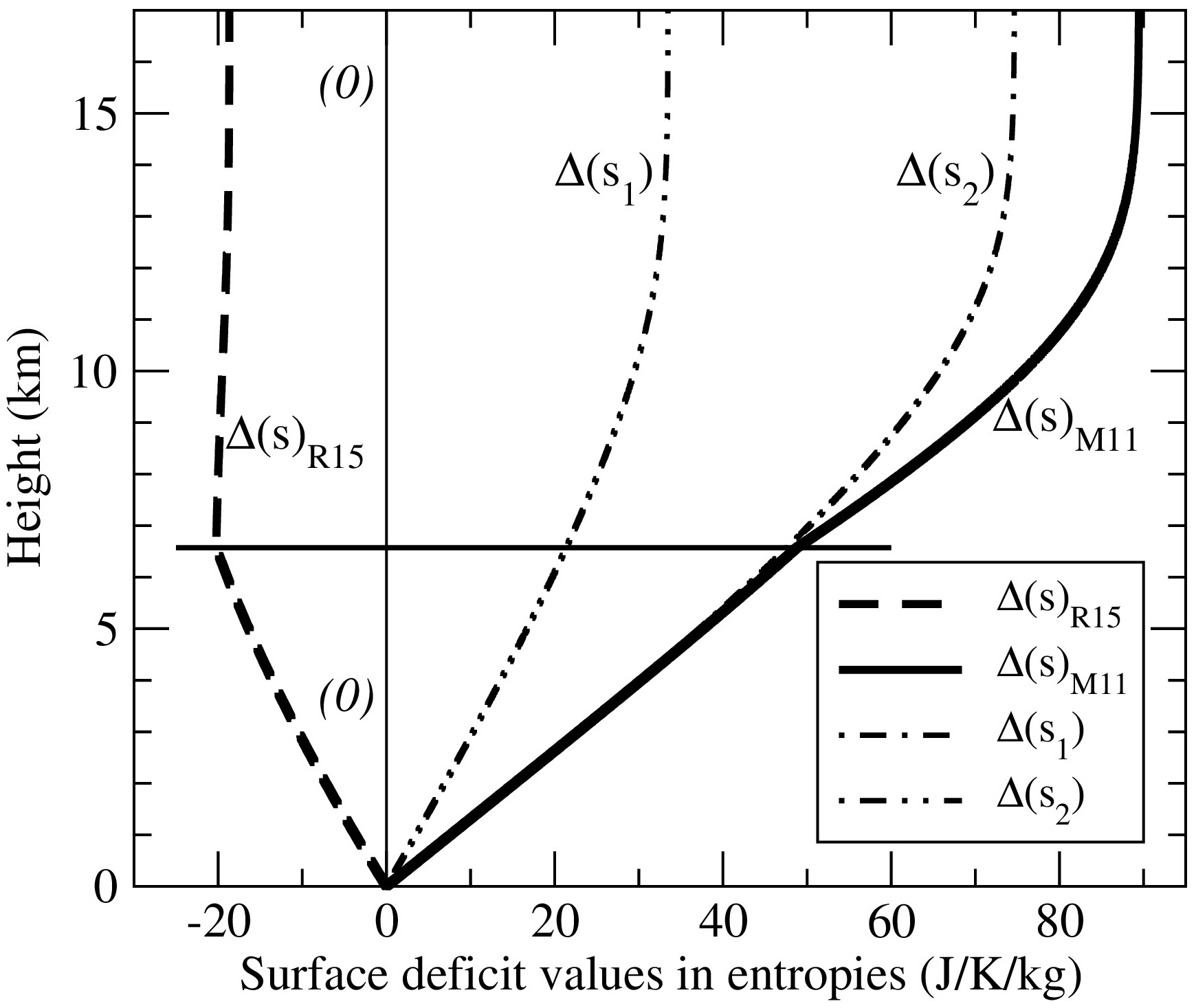}}
\vspace*{-3mm} 
\caption{\small \it 
Same as in Fig.~\ref{fig4} but for the surface deficit values in $(s)_{\rm R15}$ (dashed), $s_1$ (double-dotted dashed), $s_2$ (dotted-dashed) and $(s)_{\rm M11} = (s)_{\rm R15} + s_1 + s_2$ (solid).
Units are in J~K${}^{-1}$~kg${}^{-1}$.
} 
\label{fig5}
\end{figure}

In order to determine which entropy is correct, the surface deficit in $(s)_{\rm R15}(\theta_e)$ and $(s)_{\rm M11}(\theta_s)$ given by (\ref{eq_S_thetae_R15}) and (\ref{eq_s_theta_s}) are plotted in Fig~\ref{fig5}, together with the surface deficit of the correction terms $s_1$ and $s_2$ given by (\ref{eq_S_M11_R15_S1})-(\ref{eq_S_M11_R15_S2}).



Since the saturation value $q_v$ decreases with height for pseudo-diabatic processes, $\Delta(s_1)$ and $\Delta(s_2)$ logically increase with height because $(s_{dr}-s_{lr}) > 0$ and $\ln(p_{ar}/p_{vr}) >0$ are multiplied by the factors ``$-q_v$'' and ``$(1-q_v)$'', respectively, which both increases with height.

Moreover, the increase in $\Delta(s)_{\rm M11}(\theta_s)$ with $z$ in Fig~\ref{fig5} can be explained by using the pseudo-adiabatic change in $s$ given by Eq.(7.2) in MG15, yielding
\vspace*{-1mm}
\begin{align}
 \frac{\partial s}{\partial z}   
    & \: = \;
 \frac{c_{pd}}{\theta_s} \: \frac{\partial \theta_s}{\partial r_v}
      \; = \;
 \left( \frac{s-s_x}{1+r_x} \right)
 \left( \frac{- \: \partial r_x}{\partial z} \right)
 \label{eq_dsdz1} \: ,  \\
 \frac{\partial s}{\partial z}   
    & \: \approx \;
 (s_{dr}-s_{xr}) \: 
 \left(  \frac{R_d \: L_x}{R_v^2 \: T^2}\right)
  \left[ \: 
   \Gamma_{\rm pseudo} \;
 \left(  \frac{e_x }{p}\right)
  \: \right]
    > 0 \: . \label{eq_dsdz2}
\end{align}
The pseudo-adiabatic lapse rate $\Gamma_{\rm pseudo}$ is given by (\ref{eq_grad_pseudo_adiab}).
The terms $s_x$ and and $e_x$ are notations for the specific entropy and the saturating pressure of water with respect to liquid water for $T \geq T_{\rm trip}$ or ice for $T<T_{\rm trip}$.

The physical meaning of (\ref{eq_dsdz1}) is given by the impact of $r_x$ which decreases with $p(z)$ and $T(z)$, this creating liquid or ice precipitations, which are then withdrawn from the system.
This removal of the condensed water corresponds to the terms $(s-s_x)$, namely to a replacement of the specific quantity $s_x$ by $s \approx s_a$ in order to keep a unit mass of moist air.
The term $(s-s_x)$ can be approximated in (\ref{eq_dsdz2}) by the constant value $(s_{dr}-s_{xr})$, with good accuracy for all atmospheric values of temperature and partial pressures.

It can be checked that the approximate pseudo-adiabatic gradient (\ref{eq_dsdz2}) is roughly proportional to the bracketed term, which is almost constant with height up to $10$~km and then rapidly decreases above this level.
This explains and validates the two similar curves $\Delta\theta_s(z)$ in Fig~\ref{fig4} and $\Delta(s)_{\rm M11}(z)$ in Fig~\ref{fig5}.

Differently, the shapes of the curves $\Delta\theta_e(z)$ and $\Delta(s)_{\rm R15}(z)$ are not relevant below the freezing level, since they are both decreasing with height.
Moreover,  $\Delta(s)_{\rm R15}(z)$ is almost constant with $z$ above the freezing level, whereas $\Delta\theta_e(z)$ is more clearly increasing with height.
These differences can be explained by the varying factor $q_a = 1-q_t$ appearing both in (\ref{eq_S_thetae_R15}) and (\ref{eq_Sref_thetae_R15}), which prevents $\theta_e$ to be a true equivalent to $(s)_{\rm R15}$ for non-adiabatic processes.

The results described in this section clearly show that $\theta_e$ computed in R15 is not relevant for describing realistic profiles of moist air where $q_t$ is not a constant with height, and in particular for describing pseudo-adiabatic conditions or entrainment processes.
Clearly, the correction term $\Delta(s_1)$, which depends on $(s_{dr}-s_{lr}) =  L_{\rm vap}(T_{\rm trip})/T_{\rm trip} \: - \: c_{pd} \: \Lambda_r $, is not a small term in Fig~\ref{fig5}.
It must be taken into account in order to compute the true surface deficit in moist-air entropy $\Delta(s)_{\rm M11}$, which depends on $\theta_s$ and on $\Lambda_r = (s_{vr}-s_{dr})/c_{pd}$.
This means that it is needed to apply the third law of thermodynamics to know the reference partial entropies for  ice (Ih) and for the solid dry-air compounds at $0$~K, and then to compute the reference entropies at $T_{\rm trip}$.

 \section{Conclusion} 
\vspace*{-5mm}

It is shown in this comment that the quantity conserved in R15 is ``MSE$\: + \:$CAPE''.
The sign of the CAPE in R15 and in the title of the paper should thus be changed.

It is shown that the moist-air entropy potential temperature $\theta_s = \exp[(s-s_{ref})/c_{pd}]$ defined in M11 is an accurate alternative adiabatically conserved variable.

This comment further demonstrates that $\Delta(\mbox{MSE})_{\rm R15}$, the potential temperature $\theta_e$ and the associated moist-air entropy $(s)_{\rm R15}$ are not accurate enough for describing the realistic pseudo-adiabatic conditions or entrainment processes mentioned in the conclusion of R15.
In particular, a term depending on an arbitrary choice of reference entropy is missing.



It is demonstrated that $\theta_s$ is the only measurement of the moist-air entropy valid in all circumstances, namely for either under-saturated or saturated conditions (over liquid water or ice) and for either adiabatic or pseudo-adiabatic profiles.
It may thus be important to explain in more detail here why we must all apply the third law of thermodynamics in atmospheric science.

In fact, the main problem associated with computations of moist-air entropy has already been analyzed in  
\citet[ p.158-160]{Richardson22}.
Richardson stated that 
{\it the most natural way of reckoning the entropy of the water substance would be to take it as zero at the absolute zero of temperature\/}.

However, Richardson recalled that 
{\it it was formerly supposed that the presence of $T$ in the denominator of the integral which gives the entropy $ds = c_p(T)\: dT/T$\/}
{\it  would make the integral have an infinity where $T=0$\/}.
The advice of Richardson was to take into account the measurements of Nernst and others who showed that $c_p(T)$ of a solid tends to zero at $T=0$ in such a way that the entropy remains finite there.

This corresponds to the so-called Debye's law, which says that $c_p(T) \approx a \: T^3 $ is proportional to $T^3$ at law temperature and for all solids.
Accordingly, the equation for entropy can be written as $ds \approx a \: T^2 \: dT$, which integrates into
$s(T) \approx a \: T^3/3 + s_0$, where $s_0$ is a constant of integration.
The entropy of a solid at $T=0$~K is thus equal to $s(0) = s_0$ and $s(T$) is well-definite if $s_0$ can be determined.

Richardson added that,
{\it as there is an arbitrary constant of integration in the entropy, we must ask what would be the effect of an increase in this constant\/}, and
{\it approximations are not here permissible, for the constant might be made indefinitely large\/}.
This problem can be solved by using the third law of thermodynamics, which states that the entropy is zero for the most stable crystalline form of the substance at absolute zero temperature.
This means that $s_0 = 0$ and thus $s(T) \approx a \: T^3/3$ for all solids, with ``$a$'' a constant depending on the solid to be considered.

It is worth highlighting the advice of Richardson: the third law must not be applied to liquids or gases, only to the more stable solid state at $0$~K.
This explains why the criticisms about the third law published in Appendix~A of \citet{Pauluis_al_2010} are not valid, since they wrongly argued that the term $\ln(T)$ would be infinite at $0$~K for a perfect gas.
In fact, Debye's law is well defined and can indeed be considered for all solids, leading to finite values of entropy for all atmospheric species  (N$_2$, O$_2$, H$_2$O, Ar, CO$_2$, ...).

Richardson was not able to continue accurate computations of moist-air entropy in 1922, simply because values of $c_p(T)$ were not available at that time for all substances and for an absolute temperature varying from zero to $350$~K.
These measurements were made later, during the 1930's, for all atmospheric species and by using the magnetic refrigeration method to attain extremely low temperatures, far below $1$~K \citep{Giauque49}, thus resolving the Debye's domain close to $0$~K.

Nowadays, the third law of thermodynamics \citep{Planck17,Abriata_Laughlin_2004,Klimenko12} is considered to have been fully proved as a result of Giauque's work (see the Nobel award ceremony presentation speech by \citeauthor{Tiselius49}, \citeyear{Tiselius49}), since Giauque's measurements lead to accurate calculations of chemical affinities and to relevant predictions of the result of all chemical reactions from thermodynamic determinations of absolute entropies.

Indeed, Tiselius clearly explains that the existence or nonexistence of chemical reactions depends on the difference in free enthalpy (or Gibbs' function), with differences in entropy to be computed with values obtained from the third law, and without any other arbitrary choices such as those chosen in \citet{Romps2008} and retained in R15 or those previously chosen in \citet{Emanuel94} or \citet{Pauluis_al_2010}.







\bibliographystyle{ametsoc2014}
\bibliography{arXiv_Marquet_Romps_JAS_2015_v3}


    \end{document}